\documentclass[pdftex,singlecolumn,epjc3]{svjour3} 
\RequirePackage{graphicx}
\RequirePackage{mathptmx}      
\RequirePackage{flushend}
\RequirePackage[numbers,sort&compress]{natbib}
\RequirePackage[colorlinks,citecolor=blue,urlcolor=blue,linkcolor=blue]{hyperref}
\usepackage{amsmath}
\usepackage{graphicx}
\usepackage{geometry}
\usepackage{caption}
\usepackage{subcaption}
\usepackage{color}
\begin{document}
\baselineskip=18 pt
\begin{center}
{\large{\bf Point-like defect on Schr\"{o}dinger particles under flux field with harmonic oscillator plus Mie-type potential : application to molecular potentials}}
\end{center}

\vspace{0.1cm}

\begin{center}
{\bf Faizuddin Ahmed}\footnote{\bf faizuddinahmed15@gmail.com ; faizuddin@ustm.ac.in}\\
{\bf Department of Physics, University of Science \& Technology Meghalaya,\\
Ri-Bhoi, Meghalaya-793101, India }
\end{center}

\vspace{0.1cm}

\begin{abstract}:
In this analysis, we study the quantum motions of a non-relativistic particle confined by the Aharonov-Bohm (AB) flux field with harmonic oscillator plus Mie-type potential in a point-like defect. We determine the eigenvalue solution of the particles analytically and discuss the effects of the topological defect and flux field with this potential. This eigenvalue solution is then utilized in some diatomic molecular potential models (harmonic oscillator plus Kratzer, modified Kratzer, and attractive Coulomb potentials) and presented as the eigenvalue solutions. Afterwards, we consider a general potential form (superposition of pseudoharmonic plus Cornell-type potential) in the quantum system and analyze the effects of various factors on the eigenvalue solution. It is shown that the eigenvalue solutions get modified by the topological defect of a point-like global monopole and flux field compared to the results obtained in the flat space.
\end{abstract}

\vspace{0.75cm}


\vspace{0.5cm}

{\bf PACS Number(s):} 14.80.Hv, 03.65.-w, 03.65.Ge, 02.30.Gp, 03.65.Vf

\section{Introduction}
\label{intro}

Topology plays an important role in various physical systems in different areas of physics, such as gravitational physics, condensed matter physics and cosmology. The presence of topological defects changes the geometrical properties of space-time under consideration. These defects were formed during a phase transition in the early universe through a spontaneous symmetry-breaking mechanism as suggested in Ref. \cite{TWBK}. Topological defects are classified into cosmic strings \cite{AVV2}, Domain walls \cite{AVV}, and Global monopoles \cite{MBAV} (see also, Ref. \cite{AVV3} for detailed discussions). In condensed matter physics, the topological defect is by vortices in superconductor or super-fluid and domain walls in magnetic materials \cite{HK}, solitons in $1D$ polymers \cite{AJH}, and dislocations or disclinations in solids or liquid crystals \cite{MK}. Topology change of a medium introduced by a linear defect, such as disclination, dislocation or dispiration produces some effects on the physical properties of the medium \cite{VBB2,WCFS,WCFS2}. 

In quantum mechanical systems, the effects of the topological defect produced by a cosmic string or a spinning cosmic string space-time have been studied widely in the literature (see, Refs. \cite{ALCO,SZ2,cc2,cc3,cc4} and related references therein). Researchers have solved the wave equations without or with magnetic and quantum flux fields subject to interaction potential of various kinds and obtained the eigenvalue solutions using different methods or techniques. In addition, cosmic string space-time has also been studied in the context of the Kaluza-Klein theory in the literature (see, Refs. \cite{EPJC,SR}) and related references therein). Some other investigations of the topological defect are Dirac fermions with Kratzer-like potential under Lorentz symmetry violation in a space-time with screw dislocation \cite{SSZ}, a static composite structure under magnetic field in the spiral dislocation space-time \cite{SSZ2}, in a cosmic string space-time with distortion of a radial line into a spiral on the scalar field \cite{SSZ3}, free fermions in the presence of spiral dislocation in a space–time with distortion of a radial line into spiral \cite{SSZ4}, interactions of an electron with nonuniform electric field under cut-off point induced by spiral dislocation \cite{AVDM}, spiral dislocation on the confinement of a point charge under the rotating frame of reference \cite{WCFS3}, and interactions of an electron with magnetic field in a space-time with spiral dislocation \cite{AVDM2}. 

Another topological defect produced by a point-like global monopole (PGM) has been studied in quantum mechanical systems, for example, in the relativistic limit, quantum motions of a charged spin-$0$ particle in the presence of a dyon, Aharonov–Bohm magnetic field and scalar potential \cite{ALCO}, relativistic quantum oscillator in \cite{SZ}, with rainbow gravity in \cite{SZ4}, its generalized version in \cite{SZ3}, and with a scalar potential in \cite{SR2}. In the non-relativistic limit, these investigations are harmonic oscillator problems \cite{CF}, and with potential \cite{RV}, non-relativistic particle interacts with various potentials, such as the Kratzer and Morse potential \cite{VBB}, generalized Morse potential \cite{PN}, and diatomic molecular potential \cite{MP2}. The study of topological defects in the non-relativistic quantum system has some significance because the physical properties are changed, and the eigenvalue solution gets modified compared to the flat space results with potentials. As stated above, only a few potential models have been used to investigate the non-relativistic wave equation in the background of a point-like defect. Therefore, studies of the quantum motion of non-relativistic particles or harmonic oscillators interacting with other potentials in a point-like defect have significance in the literature which is our main motivation in this paper. Furthermore, if one introduces an electromagnetic four-vector potential in the quantum system through a minimal substitution in the wave equations, then the eigenvalue solutions of the particle get more modified in addition to the topological defect of the geometry under consideration. 

The exact or approximate eigenvalue solutions of the non-relativistic Schr\"{o}dinger equation (SE) using different techniques or methods with various physical potentials have been investigated in flat space background by many authors. These potentials include a general potential form \cite{ss1,ss3,ss9}, Mie-type potential \cite{SI,SMI5,SMI6}, Kratzer potential (KP) \cite{AK,EF,HD,AKR,HA,MRS1,OB1,KO1,SI,SMI5,SMI6,ff11,SSZ,VBB}, Morse potential \cite{PMM,VBB}, modified Kratzer potential (MKP) \cite{CB,SI,SMI5,SMI6,KO1}, highly singular potentials \cite{ff2}, hyperbolic potential \cite{ff3}, Yukawa potential \cite{HY,ff4}, generalized inversely quadratic Yukawa potential \cite{ff5}, non-central potential \cite{ff6}, modified Kratzer plus ring-shaped potential \cite{ff13}, Hulthen potential \cite{gg1,gg2}, Manning-Rosen (MR) potential \cite{gg3,gg4,gg5}, Rosen-Morse potential \cite{gg6,gg7}, Deng–Fan potential \cite{gg8}, Hyperbolic Poschl–Teller potential \cite{gg9}, pseudo-harmonic potential \cite{gg11,ss6,ss7,ss4,ss8} and many more potential well-known in the literature which has great importance in different branches of physics and chemistry. The hydrogen atom and harmonic oscillator are usually given in many textbooks as these two are several exactly solvable problems \cite{WG,LDL,FC}. According to the Schr\"{o}dinger formulation of quantum mechanics, the total wave function provides all the relevant information about the behaviour of a physical system under investigation. Hence, if it is exactly solvable for a given potential model, then the wave function can describe such a system completely.

In this analysis, We study a non-relativistic particle confined by the Aharonov-Bohm flux field in a point-like defect with potential of physical interest which is different from those potentials considered in Refs. \cite{RV,VBB}. The general form of the potential is given by \cite{ss1,ss3,ss9} 
\begin{equation}
V (r)=\beta\,r^2+\beta_1\,r+\frac{\beta_{-1}}{r}+\frac{\beta_{-2}}{r^2}+V_0,
\label{1}
\end{equation}
where one can choose $\beta=\frac{1}{2}\,M\,\omega^2$ (will be discussed in the subsequent section), $V_0$ is a constant potential term, and $\beta_{i}$, $i=-2, -1, 1$ are parameters characterise the different potential strengths. From this general potential form, one can recover some well-known potentials model (which will be discussed in the subsequent section). We solve the non-relativistic wave equation with this general potential form analytically, determine the eigenvalue solutions and analyze the effects of various factors, such as the topological defect of the geometry and the flux field on the energy levels and the wave functions. It is shown that the topological defect shifted the eigenvalue solutions and modified the results compared to the flat space with the chosen potential. This topological defect of a point-like global monopole breaks the degeneracy of the energy levels. Furthermore, the presence of the flux field in the quantum system also modifies the eigenvalue solution and shows an analogue of the Aharonov-Bohm effect. Note that the first term in the above general potential expression represents a harmonic oscillator. Alternately, one can see that we are investigating the harmonic oscillator problem in a point-like defect under the influence of the flux field with the potential of the form $\Big(\beta_1\,r+\frac{\beta_{-1}}{r}+\frac{\beta_{-2}}{r^2}+V_0\Big)$.

This paper is organised as follows: in {\it section 2}, we discuss the Schr\"{o}dinger wave equation in three-dimensions under the influence of the Aharonov-Bohm flux field in the presence of potential in a point-like global monopole. Then, we solve the radial equation with the harmonic oscillator plus Mie-type potential and obtain the eigenvalue solution analytically; in {\it section 3}, we utilized this eigenvalue solution for some diatomic molecular potential models; in {\it section 4}, we solve the radial wave equation with the potential of the general form (\ref{1}) and obtain the eigenvalue solution; in {\it section 5}, we present our results. We have used the natural units $c=1=\hbar$.

\section{Non-Relativistic Particles in Point-like Defect with Harmonic Oscillator Plus Mie-Type Potential Under AB-Flux Field }
\label{sec: 1}

We begin this section with a static and spherically symmetric space-time describing a point-like global monopole in the spherical coordinates $(t, r, \theta, \phi)$ in given by the following line-element \cite{CF,ERBM,RV,ALCO,SR2,VBB,MP2}
\begin{equation}
ds^2=-dt^2+\frac{dr^2}{\alpha^2}+r^2\,(d\theta^2+\sin^2 \theta\,d\phi^2),
\label{2}
\end{equation}
where $\alpha < 1$ represents the topological defect parameter. Here $(r, \theta, \phi)$ with $0 \leq r < \infty$, $0 \leq \theta < \pi$, and $0 \leq \phi < 2\,\pi$ are the spatial coordinates. The geometrical properties of this four-dimensional space-time of point-like global monopole were discussed in Ref. \cite{ERBM} (see, also \cite{CF,ALCO,SR2}). The spatial part of the above four-dimensional space-time can be written as $ds^2_{3D}=g_{ij}\,dx^i\,dx^j$, where $i,j=1,2,3$ and notations have used as $x^1=r, x^2=\theta, x^3=\phi$. The components of the metric tensor $g_{ij}$ with its inverse are $g_{11}=\frac{1}{\alpha^2}=\frac{1}{g^{11}}$, $g_{22}=r^2=\frac{1}{g^{22}}$, $g_{33}=r^2\,\sin^2 \theta=\frac{1}{g^{33}}$, and $g_{ij}=0$ for $i\neq j$. The determinant of this metric tensor $g_{ij}$ will be $g=|g_{ij}|=\frac{r^4\,\sin^2 \theta}{\alpha^2}$. One can see that for $\alpha \to 1$, this spatial part becomes $ds^2_{3D}=dr^2+r^2\,(d\theta^2+\sin^2 \theta\,d\phi^2)$ that we can write in the Cartesian coordinates $ds^2_{3D}=dx^2+dy^2+dz^2$. Thus, the presence of the topological defect makes the space-time geometry (\ref{2}) to be a curved space-time and its spatial part is look like a toroidal geometry with defect.

Here, the quantum motions of non-relativistic particles under the influence of the quantum flux field with potential are investigated in the background of the above point-like global monopole. Therefore, the time-dependent Schr\"{o}dinger wave equation is given by \cite{CF,RV,ALCO,WG,LDL,FC,VBB,MP2}
\begin{eqnarray}
\Bigg[-\frac{1}{2\,M}\,\Big\{\frac{1}{\sqrt{g}}\,D_{i}\,\Big(\sqrt{g}\,g^{ij}\,D_{j}\Big)\Big\}+V(r)\Bigg]\,\Psi=i\,\frac{\partial\,\Psi}{\partial\,t},
\label{4}
\end{eqnarray}
where $M$ is the particles mass, $\Psi$ is the total wave function, $D_{i} \equiv \Big(\partial_{i}-i\,e\,A_{i}\Big)$ \cite{ALCO,WG,LDL,VBB,SR2,MP2} with $e$ is the electric charges, $A_{i}$ is the electromagnetic three-vector potential, and others are mentioned earlier. 

By the method of separation of the variables, one can express the total wave function $\Psi (t, r, \theta, \phi)$ in terms of different variables. Suppose, a possible total wave function in terms of a radial wave function $\psi (r)$ is as follows:
\begin{equation}
\Psi(t, r, \theta, \phi)=e^{-i\,E\,t}\,Y_{l,m} (\theta, \phi)\,\frac{\psi (r)}{r},
\label{5}
\end{equation}
where $E$ is the particles energy, $Y_{l,m} (\theta, \phi)$ is the spherical harmonic functions, and $l, m$ are respectively the orbital and magnetic quantum numbers. In addition, we choose the following electromagnetic three-vector potential $\vec{A}$ Refs. \cite{ALCO,SR2,MP2}
\begin{equation}
A_{r}=0=A_{\theta},\quad A_{\phi}=\frac{\Phi_{AB}}{2\,\pi\,r\,\sin \theta},
\label{6}
\end{equation}
where $\Phi_{AB}=const=\Phi\,\Phi_0$ is the Aharonov-Bohm flux field, $\Phi_0=2\,\pi\,e^{-1}$ is the quantum of magnetic flux, and $\Phi$ is the amount of magnetic flux which is a positive integer.

Thereby, explicitly writing the wave equation (\ref{4}) in the space-time background (\ref{2}) and using Eqs. (\ref{5})--(\ref{6}), we obtain the following radial and angular equations:
\begin{eqnarray}
&&\psi''(r)+\frac{1}{\alpha^2}\,\Bigg[2\,M\,(E-V(r))-\frac{(l-\Phi)\,(l-\Phi+1)}{r^2}\Bigg]\,\psi (r)=0,\nonumber\\
&&-\Bigg[\frac{1}{\sin \theta}\,\frac{\partial}{\partial \theta}\,\Big(\sin \theta\,\frac{\partial}{\partial \theta}\Big)+ \frac{1}{\sin^2 \theta}\,\Big(\frac{\partial}{\partial \phi}-i\,\Phi\Big)^2\Bigg]\,Y(\theta, \phi)=l'\,(l'+1)\,Y(\theta, \phi),
\label{7}
\end{eqnarray}
where $l'=(l-\Phi)$ an effective orbital quantum number due to the flux field. Note that for zero quantum flux field $\Phi_{AB} \to 0$, one will get back the standard angular equations which were given in many textbooks \cite{WG,LDL,FC}. It's known that the angular momentum quantum number $l$ is related with the magnetic quantum number $m$ by $l=\kappa+|m|$, where $\kappa=0,1,2,...$. Thus, the flux field shifted the orbital quantum number $l$ by the same amount $\Phi$ as that of the magnetic quantum number $m(>0)$.  

From the above radial equation (\ref{7}), one can easily find the effective potential of the quantum system given by
\begin{equation}
V_{eff}=\Bigg[\frac{(l-\Phi)\,(l-\Phi+1)}{2\,M\,\alpha^2\,r^2}+\frac{V(r)}{\alpha^2} \Bigg].
\label{8}
\end{equation}
One can see that for a given potential $V(r)$, the effective potential of the quantum system depends on the topological defect of the geometry characterized by the parameter $\alpha$, and the quantum flux field $\Phi_{AB}$.

In this analysis, we consider an interesting potential the superposition of a harmonic oscillator $(\beta\,r^2)$ \cite{CF,RV} plus Mie-type potential $\Big(\frac{\beta_{-1}}{r}+\frac{\beta_{-2}}{r^2}+V_0\Big)$ \cite{SI,SMI5,SMI6} given by

\begin{figure}
\begin{subfigure}[b]{0.45\textwidth}
\includegraphics[width=2.9in, height=1.45in]{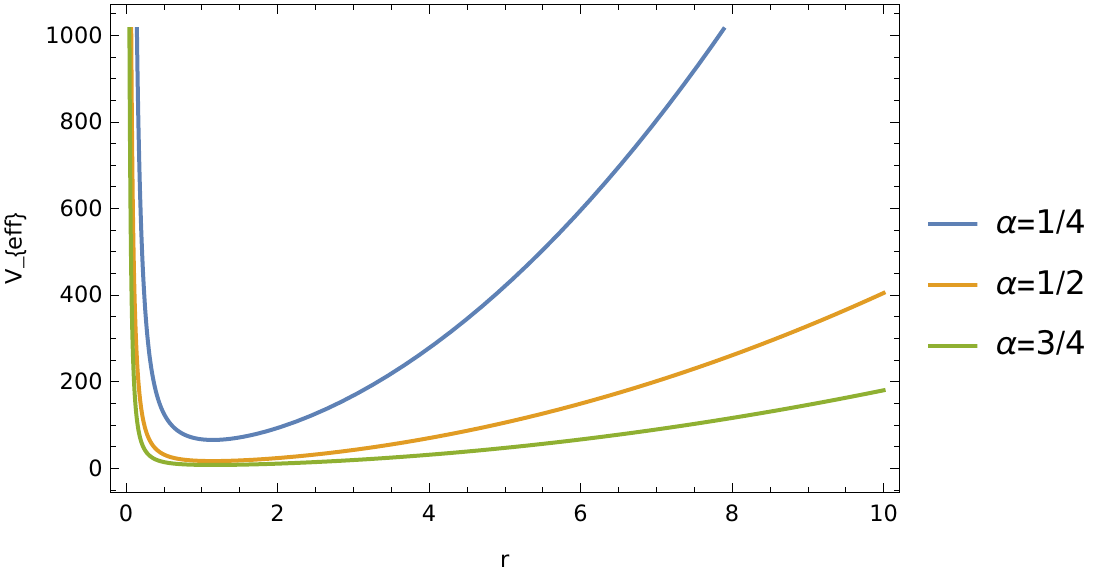}
\caption{$\Phi=3/4$, $l=1=M=\beta=\beta_{-2}=\beta_{-1}=V_0$.}
\label{fig: 1 (a)}
\end{subfigure}
\hfill
\begin{subfigure}[b]{0.45\textwidth}
\includegraphics[width=2.8in, height=1.4in]{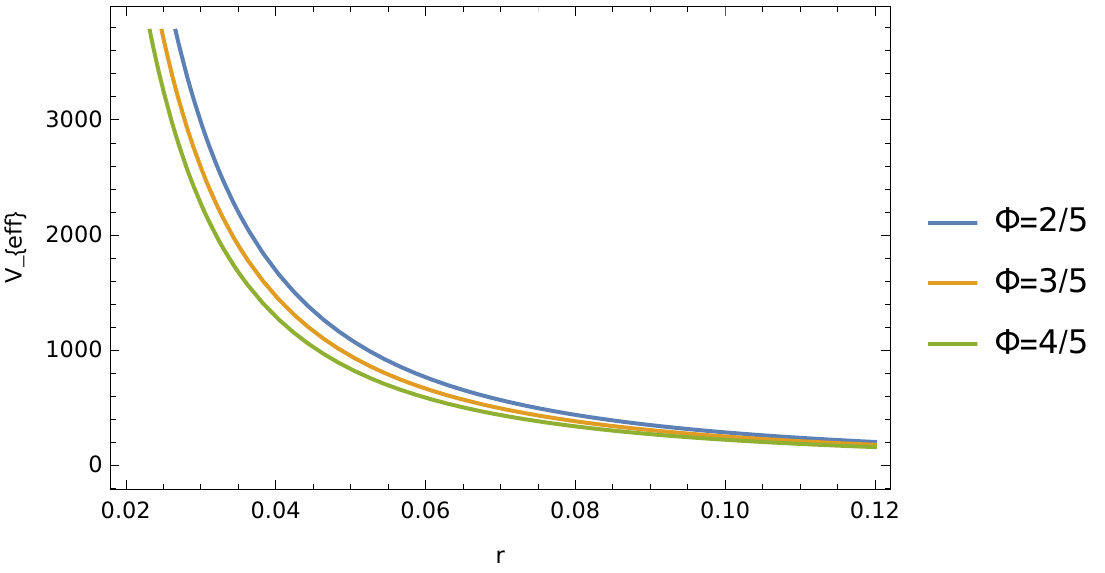}
\caption{$\alpha=3/4$, $l=1=M=\beta=\beta_{-2}=\beta_{-1}=V_0$.}
\label{fig: 1 (b)}
\end{subfigure}
\hfill\\
\begin{subfigure}[b]{0.45\textwidth}
\includegraphics[width=2.8in, height=1.4in]{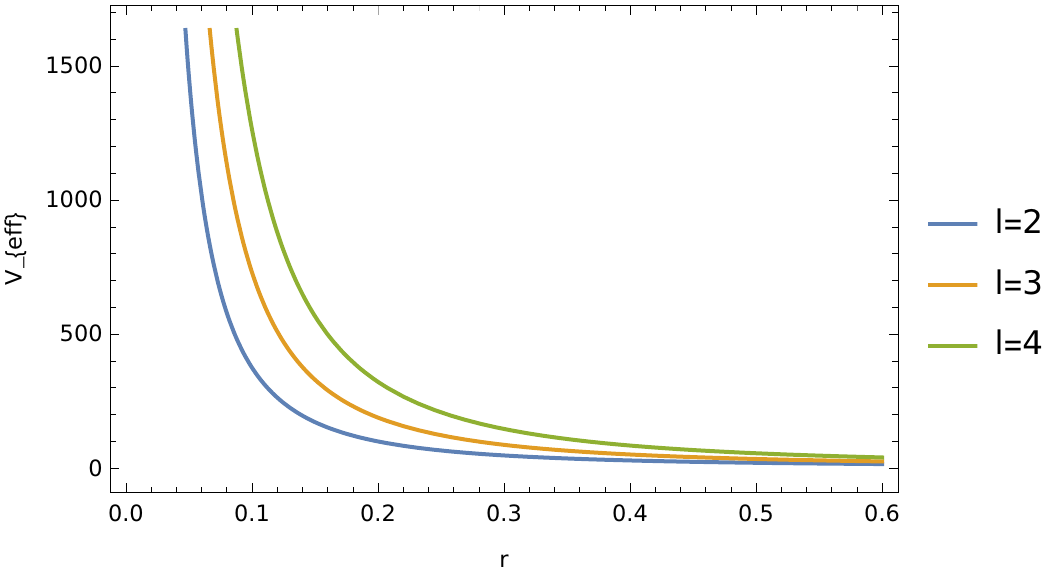}
\caption{$\alpha=3/4$, $\Phi=1$, $l=1=M=\beta=\beta_{-2}=\beta_{-1}=V_0$.}
\label{fig: 1 (c)}
\end{subfigure}
\hfill
\begin{subfigure}[b]{0.45\textwidth}
\includegraphics[width=2.8in, height=1.4in]{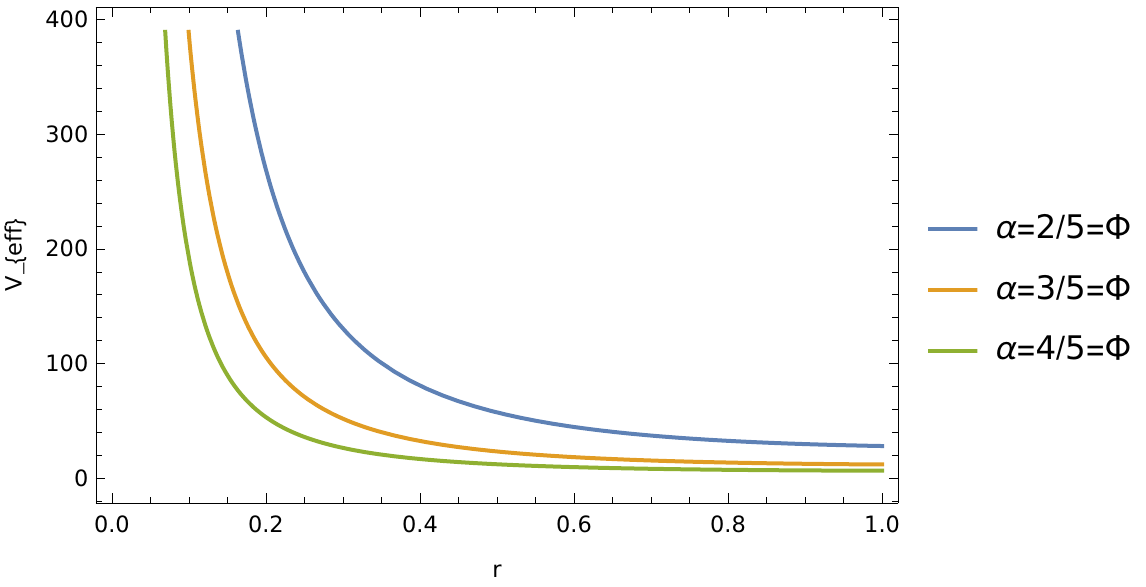}
\caption{$l=1=M=\beta=\beta_{-2}=\beta_{-1}=V_0$.}
\label{fig: 1 (d)}
\end{subfigure}
\caption{Effective potential with radial distance $r$ for different values of various parameters.}
\label{fig: 1}
\end{figure}

\begin{equation}
V (r)=\beta\,r^2+\Big(\frac{\beta_{-1}}{r}+\frac{\beta_{-2}}{r^2}+V_0\Big).
\label{bb1}
\end{equation}
One can see that for $\beta_{-1} \to 0$, one will have pseudoharmonic potential \cite{ss6,ss7,ss4,ss8}. For $\beta_{-2} \to 0$, the potential reduces to harmonic oscillator plus Coulomb-type potential that has been studied in Ref. \cite{RV} in the same space-time background. Furthermore, for $\beta \to 0$, we have Mie-type potential \cite{SI,SMI5,SMI6} from which one can recover some known molecular potentials. Using the above potential (\ref{bb1}) in the expression (\ref{8}), we have plotted a few graphs (fig. $1$) of the effective potential of the quantum system with different values of the topological defect parameter $\alpha$, the magnetic flux $\Phi$, and the orbital quantum number $l$.

Thereby, substituting potential (\ref{bb1}) in the Eq. (\ref{7}), we obtain the following differential equation of the radial function:
\begin{equation}
\psi''(r)+\Bigg(\Lambda-\gamma_2\,r^2-\frac{\gamma_{-1}}{r}-\frac{\gamma_{-2}}{r^2}\Bigg)\,\psi(r)=0,
\label{bb2}
\end{equation}
where we set the parameters
\begin{eqnarray}
\Lambda=\frac{2\,M\,(E-V_0)}{\alpha^2}\,,\,\,\,
\gamma_2=\frac{2\,M\,\beta}{\alpha^2}\,,\,\,\,
\gamma_{-1}=\frac{2\,M\,\beta_{-1}}{\alpha^2}\,,\,\,\,\gamma_{-2}=\frac{(l-\Phi)\,(l-\Phi+1)+2\,M\,\beta_{-2}}{\alpha^2}.
\label{bb3}
\end{eqnarray}

Let us perform a change of variable via $x=(\gamma_2)^{1/4}\,r$ into the Eq. (\ref{bb2}), we have obtained the following second order differential equation:
\begin{equation}
\psi''(x)+\Bigg[\Delta-x^2-\frac{\gamma_{-2}}{x^2}-\frac{\kappa}{x}\Bigg]\,\psi(x)=0,
\label{bb4}
\end{equation}
where
\begin{eqnarray}
\Delta=\frac{\Lambda}{\sqrt{\gamma_2}}\quad,\quad \kappa=\frac{\gamma_{-1}}{\gamma_{2}^{1/4}}.
\label{bb5}
\end{eqnarray}

Equation (\ref{bb4}) is the one-dimensional Schr\"{o}dinger wave equation which can be solved using different methods or techniques. Several researchers have employed or applied various methods or techniques, such as the asymptotic iteration method (AIM) \cite{ff2,ff3}, the Nikiforov-Uvarov (NU) method \cite{AFN}, supersymmetric quantum mechanics (SUSYQM) \cite{ff4,ff5,ff6}, path integral method (PIM) \cite{ff8}, factorization method \cite{ff13}, exact quantization rule \cite{ff11} and many more in order to find the exact and approximate solutions of the Schr\"{o}dinger equation. In this analysis, we approach another method where the eigenvalue solution of Eq. (\ref{bb4}) can express as the biconfluent Heun (BCH) functions. The biconfluent Heun polynomial $H (\rho, \sigma, \lambda, \mu; x)$ can obtain as a Frobenius solution to the BCH equation \cite{AR,SYS} computed as a power series expansion around the origin \cite{GBA,EPJC,SR,RV}. 

Let us choose a possible solution to the Eq. (\ref{bb4}) as follows:
\begin{equation}
\psi (x)=x^{A}\,e^{-B\,x^2}\,H (x),
\label{bb6}
\end{equation}
where $H (x)$is an unknown function.

Thereby, substituting Eq. (\ref{bb6}) into the Eq. (\ref{bb4}), we have arrived the following differential equation
\begin{eqnarray}
H''(x)+\Bigg[\frac{2\,A}{x}-4\,B\,x\Bigg]\,H'(x)+\Bigg [\frac{A^2-A-\gamma_{-2}}{x^2}-\frac{\kappa}{x}+(4\,B^2-1)\,x^2+\Big(\Delta-2\,B-4\,A\,B \Big)\Bigg]\,H (x)=0.
\label{bb7}
\end{eqnarray}
Equating the coefficients of $x^{-2}, x^{2}$ from the third term equals to zero, we have (taking positive values)
\begin{eqnarray}
&&A^2-A-\gamma_{-2}=0\Rightarrow A=\frac{1}{2}\,(1+\sqrt{1+4\,\gamma_{-2}})=\frac{1}{2}\,(1+2\,j),\quad j=\sqrt{\gamma_{-2}+\frac{1}{4}},\nonumber\\
&&4\,B^2=1\Rightarrow B=1/2.
\label{bb8}
\end{eqnarray}

Thereby, substituting $A, B$ in the Eqs. (\ref{bb6}), we obtain the radial wave function as follows
\begin{equation}
\psi (x)=x^{\frac{1}{2}}\,x^{j}\,e^{-\frac{x^2}{2}}\,H (x) \quad , \quad j=\sqrt{\frac{(l-\Phi)\,(l-\Phi+1)+2\,M\,\beta_{-2}}{\alpha^2}+\frac{1}{4}}.
\label{bb9}
\end{equation}
And that the differential equation (\ref{bb7}) becomes
\begin{equation}
H''(x)+\Bigg[\frac{1+2\,j}{x}-2\,x \Bigg]\,H'(x)+\Bigg [-\frac{\kappa}{x}+\Pi\Bigg]\,H (x)=0.
\label{bb10}
\end{equation}
where
\begin{equation}
\Pi=\Delta-2\,(1+j).
\label{bb11}
\end{equation}
Equation (\ref{bb10}) is the biconfluent Heun differential equation form \cite{AR,SYS,RV,EPJC,SR} and $H (x)$ is the Heun function.  

The differential equation (\ref{bb10}) can be solved by using the Frobenius power series solution method. Writing the power series solution around the origin \cite{GBA} by
\begin{equation}
H (x)=\sum^{\infty}_{i=0}\,d_{i}\,x^{i}.
\label{bb12}
\end{equation}
Thereby, substituting the power series solution (\ref{bb12}) in the Eq. (\ref{bb10}), one will find the following recurrence relation:
\begin{equation}
d_{n+2}=\frac{1}{(n+2)(n+2+2\,j)}\,\Big[\kappa\,d_{n+1}-(\Pi-2\,n)\,d_{n}\Big]. 
\label{bb13}
\end{equation}
With the few coefficients
\begin{equation}
d_1=\Big(\frac{\kappa}{1+2\,j}\Big)\,d_0\quad,\quad d_2=\frac{1}{4\,(1+j)}\,\Big(\kappa\,d_{1}-\Pi\,d_{0}\Big)\quad,\quad d_3=\frac{1}{6\,(j+\frac{3}{2})}\,\Big(\kappa\,d_2-2\,d_1\Big).
\label{bb14}
\end{equation}

In quantum theory, it is required that the wave function $\psi (x)$ must be well-behaved and regular everywhere for $x \to 0$ and $x \to \infty$. One can find the eigenvalue solution by imposing a condition on the Heun function $H(x)$ that it must be a finite degree polynomial of $x$ with degree $n$. Through the recurrence expression (\ref{bb8}), we can see that this power series expansion $H (x)$ becomes a polynomial of degree $n$ provided the following conditions fulfilled \cite{RV,EPJC,SR}
\begin{equation}
\Pi=2\,n\,(n=1,2,3,...)\quad,\quad d_{n+1}=0
\label{bb15}
\end{equation}
so that $d_{n+2}=0$ and the Heun function becomes a finite degree polynomial, $H (x)=(d_0+d_1\,x+d_2\,x^2+....+d_n\,x^n)$. 

After simplification of the condition $\Pi=2\,n$, we obtain the following expression of the energy eigenvalues $E_{n,l}$ given by 
\begin{equation}
E_{n,l}=V_0+\alpha\,\sqrt{\frac{2\,\beta}{M}}\,\Bigg(n+\sqrt{\frac{(l-\Phi)\,(l-\Phi+1)+2\,M\,\beta_{-2}}{\alpha^2}+\frac{1}{4}}+1\Bigg).
\label{bb16}
\end{equation}

Equation (\ref{bb16}) is the energy eigenvalue expression (non-compact) of a non-relativistic particle confined by the AB-flux field with harmonic oscillator plus Mie-type potential in a point-like defect. To have the complete information about a quantum system, one must analyze the second condition $d_{n+1}=0$ for each radial mode as done in Refs. \cite{RV,EPJC,SR}. As for example, for the radial mode $n=1$ that corresponds to the lowest state or ground state of the system, we have $\Pi=2$ and $d_2=0$ which implies from Eq. (\ref{bb14}) that 
\begin{eqnarray}
\frac{2}{\kappa}\,d_0=\Big(\frac{\kappa}{1+2\,j}\Big)\,d_0\Rightarrow \beta^{1,l}=\frac{M^3}{2\,\alpha^6}\,\Bigg(\frac{\beta^2_{-1}}{j+\frac{1}{2}}\Bigg)^2
\label{bb17}
\end{eqnarray}
a constraint on the parameter $\beta \to \beta^{1,l}$ that depends on the topological defect characterized by the parameter $\alpha$, the magnetic flux $\Phi$, and the potential parameter $\beta_{-1}$. 

Therefore, the ground state energy level is given by
\begin{equation}
E_{1,l}=V_0+\frac{M}{\alpha^2}\,\frac{(\beta_{-1})^2}{\Bigg(\sqrt{\frac{(l-\Phi)\,(l-\Phi+1)+2\,M\,\beta_{-2}}{\alpha^2}+\frac{1}{4}}+\frac{1}{2}\Bigg)}\,\Bigg(2+\sqrt{\frac{(l-\Phi)\,(l-\Phi+1)+2\,M\,\beta_{-2}}{\alpha^2}+\frac{1}{4}}\Bigg).
\label{bb18}
\end{equation}
And that the corresponding radial wave function will be
\begin{equation}
\psi_{1,l} (x)=x^{\frac{1}{2}\,(1+2\,j)}\,e^{-\frac{x^2}{2}}\,(d_0+d_1\,x)\quad,\quad d_1=\frac{1}{\sqrt{j+\frac{1}{2}}}\,d_0,
\label{bb19}
\end{equation}
where $j$ is defined in (\ref{bb9}).

Similarly, for the radial mode $n=2$, we have $\Pi=4$ and $d_3=0$ which implies from Eq. (\ref{bb14}) that
\begin{eqnarray}
d_2=\frac{2}{\kappa}\,d_0\Rightarrow \kappa=4\,\sqrt{j+\frac{3}{4}}\Rightarrow  \beta^{2,l}=\frac{M^3}{32\,\alpha^6}\,\Bigg(\frac{\beta^2_{-1}}{j+\frac{3}{4}}\Bigg)^2
\label{bb20}
\end{eqnarray}
another constraint on the parameter $\beta \to \beta^{2,l}$. Thus, we can see that for each radial mode, we have a different relation of the potential parameter $\beta \to \beta^{n,l}$ that depends on the topological defect, the magnetic flux, and the potential parameter $\beta_{-1}$.

Therefore, the first excited state energy level for the radial mode $n=2$ is given by 
\begin{equation}
E_{2,l}=V_0+\frac{M}{4\,\alpha^2}\,\frac{(\beta_{-1})^2}{\Bigg(\sqrt{\frac{(l-\Phi)\,(l-\Phi+1)+2\,M\,\beta_{-2}}{\alpha^2}+\frac{1}{4}}+\frac{3}{4}\Bigg)}\,\Bigg(3+\sqrt{\frac{(l-\Phi)\,(l-\Phi+1)+2\,M\,\beta_{-2}}{\alpha^2}+\frac{1}{4}}\Bigg).
\label{bb21}
\end{equation}
And the corresponding radial wave function will be 
\begin{equation}
\psi_{2,l} (x)=x^{\frac{1}{2}\,(1+2\,j)}\,e^{-\frac{x^2}{2}}\,(d_0+d_1\,x+d_2\,x^2),
\label{bb22}
\end{equation}
where the coefficients are
\begin{equation}
d_1=2\,\frac{\sqrt{j+\frac{3}{4}}}{\Big(j+\frac{1}{2}\Big)}\,d_0\quad,\quad d_2=\frac{1}{\Big(j+\frac{1}{2}\Big)}\,d_0.
\label{bb23}
\end{equation}

We can see that the energy levels $E_{1,l}, E_{2,l}, E_{3,l},....$ and the radial wave functions $\psi_{1,l}, \psi_{2,l}, \psi_{3,l},..$ of non-relativistic particles are influenced by the topological defect of a point-like global monopole characterized by the parameter $\alpha$ and shifted the eigenvalue solution compared to the flat space result with this superposed potential. Furthermore, we see that the quantum flux field $\Phi_{AB}$ shifted the eigenvalue solution more in addition to the topological defect and shows an analogue of the Aharonov-Bohm effect \cite{YA,MP} because the energy eigenvalue depends on the geometric quantum phase.

\section{ APPLICATIONS TO DIATOMIC MOLECULAR POTENTIALS }

The above eigenvalue solution of the quantum system is now being utilize to develop solutions of the particles to some specific types of interacting molecular potentials which have wide application in practical problems.

\subsection{\bf HARMONIC OSCILLATOR PLUS KRATZER POTENTIAL}

The harmonic oscillator plus Kratzer potential can be recovered by setting the parameters $\beta=\frac{1}{2}\,M\,\omega^2$, $\beta_{-1}=-2\,D_e\,r_0$, $\beta_{-2}=D_e\,r^2_{0}$, and $V_0=0$ in the potential expression (\ref{bb1}), we obtain the following potential form
\begin{equation}
V (r)=\frac{1}{2}\,M\,\omega^2\,r^2-\frac{2\,D_e\,r_0}{r}+\frac{D_e\,r^2_{0}}{r^2}.
\label{cc1}
\end{equation}
The above potential can be written as
\begin{equation}
V (r)=\frac{1}{2}\,M\,\omega^2\,r^2+2\,D_e\,\Bigg[\frac{1}{2}\,\Big(\frac{r_0}{r}\Big)^2-\frac{r_0}{r}\Bigg].
\label{cc2}
\end{equation}
Here $\omega$ is the oscillator frequency, $D_e$ is the dissociation energy between two atoms in a solid, $r_0$ is the equilibrium internuclear separation. The second part of the above potential called Kratzer potential has of great importance in molecular physics and quantum chemistry \cite{AK,EF}, in internuclear vibration of diatomic molecules \cite{HD,AKR} and other branches of physics and chemistry \cite{HA,MRS1,OB1,KO1,SI,SMI5,SMI6}.

Thus, using the above potential Eq. (\ref{cc2}) in the radial Eq. (\ref{7}) and following the previous procedure, one will find the following energy expression  
\begin{equation}
E_{n,l}=\alpha\,\omega_{n,l}\,\Bigg(n+1+\sqrt{\frac{(l-\Phi)\,(l-\Phi+1)+2\,M\,D_e\,r^2_{0}}{\alpha^2}+\frac{1}{4}}\Bigg),
\label{cc3}
\end{equation}
where $\omega \to \omega_{n,l}$ that is clear from the above analysis. Equation (\ref{cc3}) is the non-relativistic energy eigenvalue expression of the particles confined by the Aharonov-Bohm flux field with harmonic oscillator plus Kratzer potential in a point-like defect. In other words, equation (\ref{cc3}) is the non-compact energy expression of a harmonic oscillator under the influences of the quantum flux field with Kratzer potential in the background of a point-like global monopole. 

As done earlier, one can evaluate the individual energy levels and radial wave functions one by one. As special case, for the radial mode $n=1$, one will find the following relation
\begin{eqnarray}
\omega_{1,l}=\Bigg(\frac{4\,M\,D^2_{e}\,r^2_{0}}{\alpha^3}\Bigg)\,\frac{1}{\Big(\varsigma+\frac{1}{2}\Big)},
\label{cc4}
\end{eqnarray}
a constraint on the oscillator frequency $\omega_{1,l}$ that depends on the topological defect characterized by the parameter $\alpha$, and the magnetic flux $\Phi$. It's value change with the orbital quantum number $l$. 

The ground state energy level and the radial wave function will be 
\begin{eqnarray}
E_{1,l}=\Bigg(\frac{4\,M\,D^2_{e}\,r^2_{0}}{\alpha^2}\Bigg)\,\frac{(\varsigma+2)}{\Big(\varsigma+\frac{1}{2}\Big)}\quad,\quad 
\psi_{1,l} (x)=x^{\frac{1}{2}+\varsigma}\,e^{-\frac{x^2}{2}}\,(d_0+d_1\,x),
\label{cc5}
\end{eqnarray}
where
\begin{equation}
d_1=\frac{1}{\sqrt{\varsigma+\frac{1}{2}}}\,d_0,\quad \varsigma=\sqrt{\frac{(l-\Phi)\,(l-\Phi+1)+2\,M\,D_e\,r^2_{0}}{\alpha^2}+\frac{1}{4}}.
\label{cc8}
\end{equation}

Similarly, for the radial quantum number $n=2$, we have another constraint on the oscillator frequency $\omega \to \omega_{2,l}$ given by
\begin{eqnarray}
\omega_{2,l}=\frac{M\,D^2_{e}\,r^2_{0}}{\alpha^3\,\Big(\varsigma+\frac{3}{4}\Big)}.
\label{cc6}
\end{eqnarray}

The energy level and radial wave function for the radial mode $n=2$ are given by 
\begin{eqnarray}
E_{2,l}&=&\Bigg(\frac{M\,D^2_{e}\,r^2_{0}}{\alpha^2}\Bigg)\,\frac{(\varsigma+3)}{\Big(\varsigma+\frac{3}{4}\Big)}\quad,\quad  
\psi_{2,l}(x)=x^{\frac{1}{2}\,(1+2\,\varsigma)}\,e^{-\frac{x^2}{2}}\,(d_0+d_1\,x+d_2\,x^2),\nonumber\\ 
d_1&=&2\,\frac{\sqrt{\varsigma+\frac{3}{4}}}{\Big(\varsigma+\frac{1}{2}\Big)}\,d_0,\quad d_2=\frac{1}{\Big(\varsigma+\frac{1}{2}\Big)}\,d_0,
\label{cc7}
\end{eqnarray}
where $\varsigma$ is given in (\ref{cc8}).

Thus, we can see that the energy levels $E_{1,l}, E_{2,l},...$ and radial wave function $\psi_{1,l}, \psi_{2,l},...$ of a harmonic oscillator are influenced by the topological defect of a point-like global monopole characterized by the parameter $\alpha$, and the flux field $\Phi_{AB}$ with the Kratzer potential and get them modified compared to the flat space result.

\subsection{\bf  HARMONIC OSCILLATOR PLUS MODIFIED KRATZER or KRATZER-FUES POTENTIAL}

The harmonic oscillator plus modified Kratzer or Kratzer-Fues potential can be recovered by setting the parameters $\beta_2=\frac{1}{2}\,M\,\omega^2$, $\beta_{-1}=-2\,D_e\,r_0$, $\beta_{-2}=D_e\,r^2_{0}$, and $V_0=D_e$ in the potential expression (\ref{bb1}), we obtain the following form
\begin{equation}
V (r)=\frac{1}{2}\,M\,\omega^2\,r^2-\frac{2\,D_e\,r_0}{r}+\frac{D_e\,r^2_{0}}{r^2}+D_e.
\label{dd1}
\end{equation}
That may be written as
\begin{equation}
V (r)=\frac{1}{2}\,M\,\omega^2\,r^2+D_e\,\Bigg(\frac{r-r_0}{r}\Bigg)^2.
\label{dd2}
\end{equation}
The second term in the above potential is called the modified Kratzer or Kratzer-Fues potential and has been used by several authors in the literature \cite{CB,SI,SMI6,KO1}.

Therefore, using the above potential (\ref{dd2}) in the radial Eq. (\ref{7}) and following the previous procedure, one will find the following energy eigenvalue expression  
\begin{equation}
E_{n,l}=D_e+\alpha\,\omega_{n,l}\,\Bigg(n+1+\sqrt{\frac{(l-\Phi)\,(l-\Phi+1)+2\,M\,D_e\,r^2_{0}}{\alpha^2}+\frac{1}{4}}\Bigg),
\label{dd3}
\end{equation}
Equation (\ref{dd3}) is the energy spectra of a harmonic oscillator confined by the Aharonov-Bohm flux field with modified Kratzer potential in a point-like global monopole. 

Here also, one can evaluate the individual energy level and radial wave function as done earlier. The ground state energy level $E_{1,l}$ and the radial wave function $\psi_{1,l}$ for the radial mode $n=1$ are given by
\begin{eqnarray}
E_{1,l}=D_e+\Bigg(\frac{4\,M\,D^2_{e}\,r^2_{0}}{\alpha^2}\Bigg)\,\frac{(2+\varsigma)}{\Big(\varsigma+\frac{1}{2}\Big)},\quad 
\psi_{1,l}(x)=x^{\frac{1}{2}\,(1+2\,\varsigma)}\,e^{-\frac{x^2}{2}}\,(d_0+d_1\,x),\quad  d_1=\frac{1}{\sqrt{\varsigma+\frac{1}{2}}}\,d_0,
\label{dd4}
\end{eqnarray}
where $\varsigma$ is given by (\ref{cc8}).

Similarly, for the radial mode $n=2$ the energy level $E_{2,l}$ and radial wave function $\psi_{2,l}$ are given by 
\begin{eqnarray}
E_{2,l}&=&D_e+\Bigg(\frac{M\,D^2_{e}\,r^2_{0}}{\alpha^2}\Bigg)\,\frac{(\varsigma+3)}{\Big(\varsigma+\frac{3}{4}\Big)},\quad 
\psi_{2,l}(x)=x^{\frac{1}{2}\,(1+2\,\varsigma)}\,e^{-\frac{x^2}{2}}\,(d_0+d_1\,x+d_2\,x^2),\nonumber\\
d_1&=&2\,\frac{\sqrt{\varsigma+\frac{3}{4}}}{\Big(\varsigma+\frac{1}{2}\Big)}\,d_0,\quad d_2=\frac{1}{\Big(\varsigma+\frac{1}{2}\Big)}\,d_0.
\label{dd5}
\end{eqnarray}

Thus, one can see that the energy levels $E_{1,l}, E_{2,l},.....$ and radial wave function $\psi_{1,l}, \psi_{2,l},.....$ of a harmonic oscillator are influenced by the topological defect of a point-like global monopole characterized by the parameter $\alpha$, and the flux field $\Phi_{AB}$ with modified Kratzer potential and get them modified compared to the flat space results.

\subsection{\bf HARMONIC OSCILLATOR PLUS COULOMB POTENTIAL}

The harmonic oscillator with attractive Coulomb potential can be recovered from the potential (\ref{bb1}) by setting the parameters $\beta_2=\frac{1}{2}\,M\,\omega^2$, $\beta_{-1}=-\eta_{c}$, $\beta_{-2}=0$, and $V_0=0$. Thus, we obtain
\begin{equation}
V (r)=\frac{1}{2}\,M\,\omega^2\,r^2+\Big(-\frac{\eta_c}{r}\Big).
\label{ff1}
\end{equation}

Thereby, substituting this potential (\ref{ff1}) in the radial Eq. (\ref{7}) and following the same procedure, one will find the following energy eigenvalue expression 
\begin{equation}
E_{n,l}=\omega_{n,l}\,\Bigg[(n+1)\,\alpha+\sqrt{(l-\Phi)\,(l-\Phi+1)+\frac{\alpha^2}{4}}\Bigg].
\label{ff2}
\end{equation}
The corresponding radial wave function will be
\begin{equation}
\psi_{n,l}(x)=x^{\frac{1}{2}}\,x^{\tau}\,e^{-\frac{x^2}{2}}\,H (x) \quad , \quad \tau=\sqrt{\frac{(l-\Phi)\,(l-\Phi+1)}{\alpha^2}+\frac{1}{4}}.
\label{ff3}
\end{equation}

As special cases, for the radial mode $n=1$, one will find a constraint on the oscillator frequency given by
\begin{equation}
\omega_{1,l}=\frac{M\,\eta^2_{c}}{\alpha^3\,\Big(\tau+\frac{1}{2}\Big)}.
\label{ff6}
\end{equation}
The ground state eigenvalue solution is given by
\begin{eqnarray}
E_{1,l}&=&\omega_{1,l}\,\Bigg[2\,\alpha+\sqrt{(l-\Phi)\,(l-\Phi+1)+\frac{\alpha^2}{4}} \Bigg],\nonumber\\
\psi_{1,l}(x)&=&x^{\frac{1}{2}\,(1+2\,\tau)}\,e^{-\frac{x^2}{2}}\,(d_0+d_1\,x)\quad,\quad d_1=\frac{1}{\sqrt{\tau+\frac{1}{2}}}\,d_0,
\label{ff4}
\end{eqnarray}
where $\tau$ is given in Eq. (\ref{ff3}) and $\omega_{1,l}$ in (\ref{ff6}).

Similarly, for the radial mode $n=2$, we have another constraint on the oscillator frequency given by
\begin{equation}
\omega_{2,l}=\frac{M\,\eta^2_{c}}{4\,\alpha^3\,\Big(\tau+\frac{3}{4}\Big)}.
\label{ff7}
\end{equation}

The energy level $E_{2,l}$ and the radial wave function $\psi_{2,l}$ for the radial mode $n=2$ are given by 
\begin{eqnarray}
E_{2,l}&=&\omega_{2,l}\,\Bigg[3\,\alpha+\sqrt{(l-\Phi)\,(l-\Phi+1)+\frac{\alpha^2}{4}}\Bigg],\quad 
\psi_{2,l}(x)=x^{\frac{1}{2}\,(1+2\,\tau)}\,e^{-\frac{x^2}{2}}\,(d_0+d_1\,x+d_2\,x^2),\nonumber\\
d_1&=&2\,\frac{\sqrt{\tau+\frac{3}{4}}}{\Big(\tau+\frac{1}{2}\Big)}\,d_0\quad,\quad d_2=\frac{1}{\Big(\tau+\frac{1}{2}\Big)}\,d_0,
\label{ff5}
\end{eqnarray}
where $\tau$ is given in Eq. (\ref{ff3}) and $\omega_{2,l}$ in (\ref{ff7}).

One can see that the energy levels $E_{1,l}, E_{2,l},....$ and radial wave function $\psi_{1,l}, \psi_{2,l},....$ of a harmonic oscillator are influenced by the topological defect of a point-like global monopole and the flux field in presence of an attractive Coulomb potential. 

It is worth mentioning that for zero magnetic flux field $\Phi_{AB} \to 0$, the energy levels $E_{1,l}, E_{2,l},..$ and the radial wave function $\psi_{1,l}, \psi_{2,l},...$ reduces to the result obtained in Ref. \cite{RV}. Thus, we can see that the presence of magnetic flux $\Phi$ in the quantum system modified the eigenvalue solution compared to the result obtained in Ref. \cite{RV} in the same geometry background which shows an analogue of the AB-effect \cite{YA,MP}.

\begin{figure}
\begin{subfigure}[b]{0.45\textwidth}
\includegraphics[width=2.8in, height=1.4in]{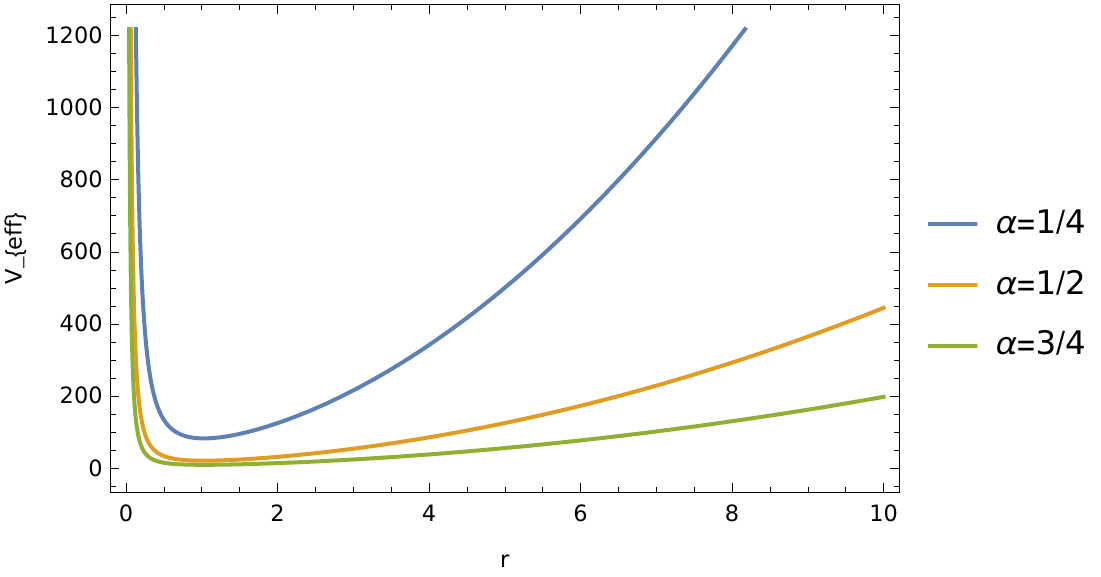}
\caption{$\Phi=3/4$, $l=1=M=\beta=\beta_{-2}=\beta_{-1}=\beta_1=V_0$.}
\label{fig: 2 (a)}
\end{subfigure}
\hfill
\begin{subfigure}[b]{0.45\textwidth}
\quad\includegraphics[width=2.8in, height=1.4in]{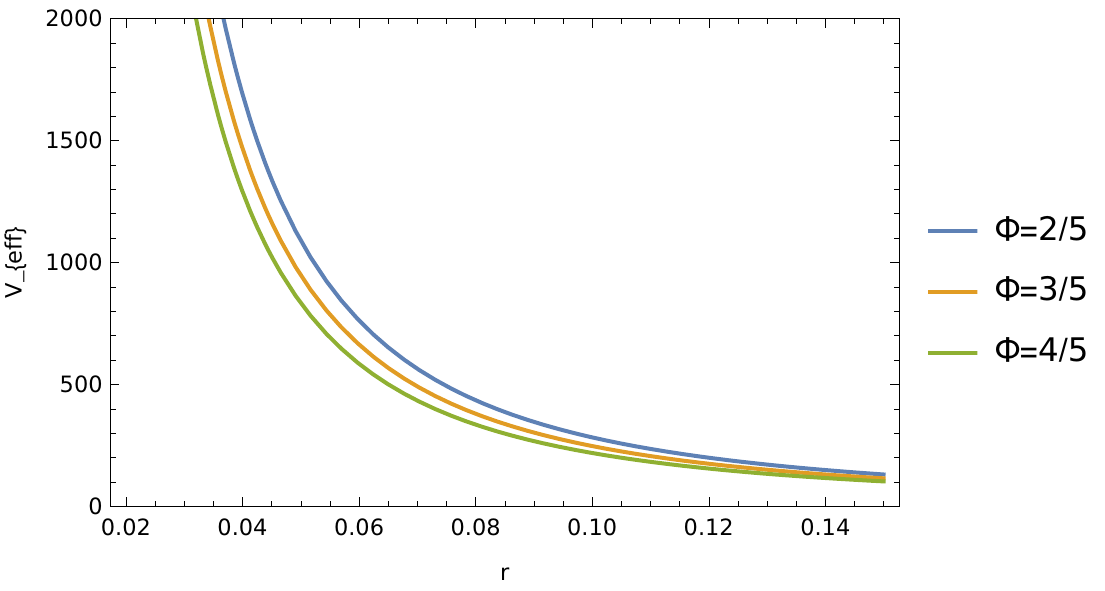}
\caption{$\alpha=3/4$, $l=1=M=\beta=\beta_{-2}=\beta_{-1}=\beta_1=V_0$.}
\label{fig: 2 (b)}
\end{subfigure}
\hfill\\
\begin{subfigure}[b]{0.45\textwidth}
\includegraphics[width=2.7in, height=1.4in]{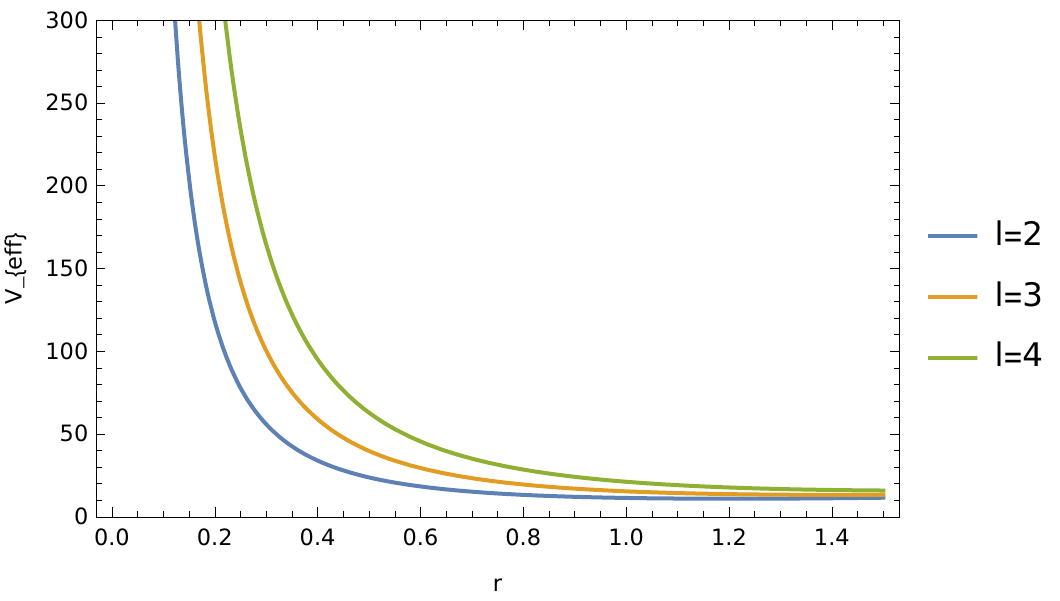}
\caption{$\alpha=3/4=\Phi$, $M=1=\beta=\beta_{-2}=\beta_{-1}=\beta_1=V_0$.}
\label{fig: 2 (c)}
\end{subfigure}
\hfill
\begin{subfigure}[b]{0.45\textwidth}
\quad\includegraphics[width=2.8in, height=1.4in]{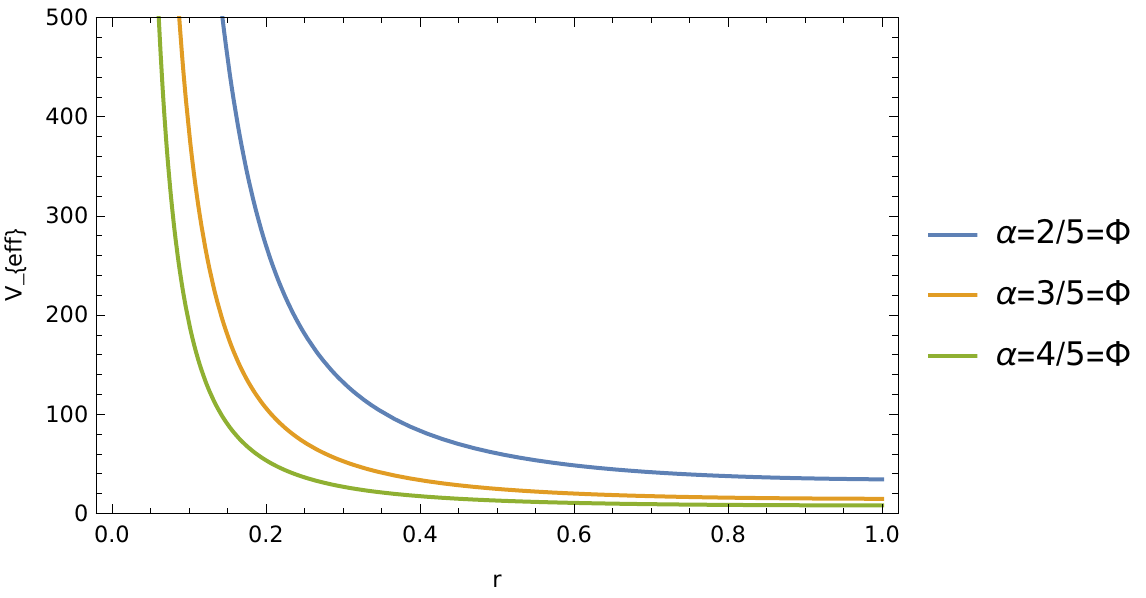}
\caption{$l=1=M=\beta=\beta_{-2}=\beta_{-1}=\beta_1=V_0$.}
\label{fig: 2 (d)}
\end{subfigure}
\caption{Effective potential with radial distance $r$ for different values of various parameters.}
\label{fig: 2}
\end{figure}

\section{ Non-Relativistic Particles Under AB-flux field in a Point-like Defect with Pseudoharmonic plus Cornell-type Potential}

In this section, we will study the quantum motions of non-relativistic particles in the presence of the AB-flux field in a point-like defect with a potential (\ref{1}) given by
\begin{equation}\nonumber
V (r)=\beta\,r^2+\beta_1\,r+\frac{\beta_{-1}}{r}+\frac{\beta_{-2}}{r^2}+V_0,
\label{9}
\end{equation}
where $\beta_1$ is a parameter characterise the linear confining potential and others are mentioned earlier. We can see that this general form of potential is the superposition of pseudo-harmonic plus Cornell-type potential or harmonic oscillator potential and inverse quadratic plus Cornell-type potential. For $\beta_{-1} \to 0$ and $\beta_1 \to 0$, the above potential reduces to a pseudo-harmonic potential \cite{ss6,ss7,ss4,ss8}. Furthermore, for $\beta_{-2} \to 0$, the potential reduces to a harmonic oscillator plus Cornell-type potential that has been studied in Ref. \cite{RV}. Using the above potential (\ref{9}) in (\ref{8}), we have plotted few graphs of the effective potential of the quantum system showing the influences of various factors, such as topological defects characterise by the parameter $\alpha$, non-minimal coupling parameter $\xi$, and the magnetic flux $\Phi$ (fig. 2).

Thereby, substituting potential (\ref{1}) in the Eq. (\ref{7}), we obtain the following radial wave equation:
\begin{equation}
\psi''(r)+\Bigg(\Lambda-\gamma_1\,r-\gamma_2\,r^2-\frac{\gamma_{-1}}{r}-\frac{\gamma_{-2}}{r^2}\Bigg)\,\psi(r)=0,
\label{10}
\end{equation}
where we have defined 
\begin{eqnarray}
\Lambda&=&\frac{2\,M\,(E-V_0)}{\alpha^2}\quad,\quad \gamma_1=\frac{2\,M\,\beta_1}{\alpha^2}\quad,\quad \gamma_{-1}=\frac{2\,M\,\beta_{-1}}{\alpha^2}\quad,\quad \gamma_2=\frac{2\,M\,\beta}{\alpha^2},\nonumber\\
\gamma_{-2}&=&\frac{(l-\Phi)\,(l-\Phi+1)+2\,M\,\beta_{-2}}{\alpha^2}.
\label{11}
\end{eqnarray}

Let us perform a change of variable via $x=(\gamma_2)^{1/4}\,r$ in the Eq. (\ref{10}), we obtain the following second-order differential equation:
\begin{equation}
\psi''(x)+\Big[\Delta-\chi\,x-x^2-\frac{\gamma_{-2}}{x^2}-\frac{\kappa}{x}\Big]\,\psi(x)=0,
\label{12}
\end{equation}
where we have set the parameters
\begin{eqnarray}
\Delta=\frac{\Lambda}{\sqrt{\gamma_2}}\quad,\quad \chi=\frac{\gamma_1}{\gamma_{2}^{3/4}}\quad,\quad \kappa=\frac{\gamma_{-1}}{\gamma_{2}^{1/4}}.
\label{13}
\end{eqnarray}
Equation (\ref{20}) is the one-dimensional Schr\"{o}dinger wave equation. As stated earlier, we can solve this equation using the BCH procedure. Let us choose a possible solution to the Eq. (\ref{12}) as follows:
\begin{equation}
\psi (x)=x^{A}\,e^{-(B\,x^2+C\,x)}\,H (x),
\label{14}
\end{equation}
where $H (x)$is an unknown function.

Substituting Eq. (\ref{14}) into the Eq. (\ref{12}), we have arrived the following equation
\begin{eqnarray}
&&H''(x)+\Bigg[\frac{2\,A}{x}-2\,C-4\,B\,x\Bigg]\,H'(x)+\Bigg[\frac{A^2-A-\gamma_{-2}}{x^2}-\frac{2\,A\,C+\kappa}{x}+(4\,B\,C-\chi)\,x\nonumber\\
&&+(4\,B^2-1)\,x^2+(\Delta-2\,B-4\,A\,B+C^2)\Bigg]\,H(x)=0.
\label{15}
\end{eqnarray}
Equating the coefficients of $x^{-2}, x^{1}, x^{2}$ from the third term equals to zero, we have
\begin{eqnarray}
&&A^2-A-\gamma_{-2}=0\Rightarrow A=\frac{1}{2}\,(1+\sqrt{1+4\,\gamma_{-2}})=\frac{1}{2}\,(1+2\,j)\quad,\quad j=\sqrt{\gamma_{-2}+\frac{1}{4}},\nonumber\\
&&4\,B^2=1\Rightarrow B=1/2\quad \mbox{and}\quad 4\,B\,C=\chi \Rightarrow C=\frac{\chi}{2}.
\label{16}
\end{eqnarray}

Substituting $A, B, C$ in the Eq. (\ref{15}), one will arrive at the following differential equation
\begin{equation}
H''(x)+\Big[\frac{1+2\,j}{x}-2\,x-\chi\Big]\,H'(x)+\Big[-\frac{\zeta}{x}+\sum\Big]\,H (x)=0,
\label{17}
\end{equation}
where we have defined
\begin{equation}
\zeta=\kappa+\frac{\chi}{2}\,(1+2\,j)\quad,\quad \sum=\Delta+\frac{\chi^2}{4}-2\,(1+j).
\label{18}
\end{equation}
Equation (\ref{17}) is the biconfluent Heun’s differential equation form \cite{SYS,AR,RV,EPJC,SR} and $H (x)$ is the Heun function. 

The radial wave function is given by
\begin{equation}
\psi (x)=x^{\frac{1}{2}\,(1+2\,j)}\,e^{-\frac{1}{2}\,(x+\chi)\,x}\,H (x),
\label{19}
\end{equation}
where $j$ is defined in Eq. (\ref{bb9}).

As stated earlier, to obtain solution of the quantum system, substituting the power series solution (\ref{bb12}) in the Eq. (\ref{17}), one will find a recurrence relation of the following form
\begin{equation}
d_{n+2}=\frac{1}{(n+2)(n+2+2\,j)}\,\Bigg[\Big\{\kappa+\chi\,\left(n+j+\frac{3}{2}\right)\Big\}\,d_{n+1} -\Big(\sum-2\,n\Big)\,d_{n}\Bigg]. 
\label{20}
\end{equation}

As stated earlier, the wave-function $\psi (x)$ must be well-behaved for $x \to 0$ and $x \to \infty$. One can obtain the bound-state solutions by imposing a condition on the Heun function $H(x)$ that it must be a finite degree polynomial of degree $n$. Through the expression (\ref{20}), one can see that the power series expansion $H (x)$ becomes a polynomial of degree $n$ provided \cite{RV,EPJC,SR} we have
\begin{equation}
\sum=2\,n\,(n=1,2,3,...)\quad,\quad d_{n+1}=0.
\label{21}
\end{equation}

After simplification of the first condition $\sum=2\,n$, we obtain the following expression of the energy $E_{n,l}$ given by 
\begin{equation}
E_{n,l}=V_0+\alpha\,\sqrt{\frac{2\,\beta}{M}}\,\Bigg(n+1+\sqrt{\frac{(l-\Phi)\,(l-\Phi+1)+2\,M\,\beta_{-2}}{\alpha^2}+\frac{1}{4}}\Bigg)-\frac{\beta^2_{1}}{4\,\beta}.
\label{22}
\end{equation}
And that the radial wave function will be
\begin{equation}
\psi_{n,l} (x)=x^{\frac{1}{2}\,(1+2\,j)}\,e^{-\frac{1}{2}\,\Big[x+\Big(\frac{2\,M\,\beta^{4}_{1}}{\alpha^2\,\beta^3}\Big)^{1/4}\Big]\,x}\,H (x),
\label{23}
\end{equation}
where $j=\sqrt{\frac{(l-\Phi)\,(l-\Phi+1)+2\,M\,\beta_{-2}}{\alpha^2}+\frac{1}{4}}$.

Equation (\ref{22}) is the non-compact expression of the energy profile of non-relativistic particles in the presence of the Aharonov-Bohm flux field with a potential of the general form (\ref{9}) in a point-like defect. We can see that the energy eigenvalue Eq. (\ref{22}) gets modified compared to the result obtained in Ref. \cite{RV} due to the presence of an inverse quadratic potential ($\sim \frac{1}{r^2}$), and the flux field $\Phi_{AB}$ which shows an analogue of the Aharonov-Bohm effect \cite{YA,MP}.

\section{Conclusions}

In the literature, all the investigations of the non-relativistic wave equation with different physical potentials have been investigated in the flat space background except those in Refs. \cite{CF,RV,VBB,PN,MP2}. They have obtained the eigenvalue solutions of the wave equation using various techniques stated in the introduction. In Refs. \cite{CF,RV,VBB,PN,MP2}, authors have studied the non-relativistic wave equation with a few known potentials in the topological defect geometry and obtained the eigenvalue solutions. They have shown that the presence of the topological defect breaks the degeneracy of the energy levels and shifts the results compared to the flat space. In this paper, we have studied another work on the effects of the topological defect on the particles confined by the Aharonov-Bohm flux field in the presence of potential of various kinds other than previous works. We have verified that the global feature of the point-like global monopole characterized by the parameter $\alpha <1$ is present explicitly in the energy levels and wave functions and shifted the eigenvalue solutions compared to the flat space results with the chosen potential. We have seen that the topological defect of a point-like global monopole breaks the degeneracy of the non-relativistic energy levels similar to those investigations done in Refs. \cite{CF,RV,VBB,PN,MP2} but with different potentials. In addition, the eigenvalue solutions get modified by the quantum flux field and the chosen potential. We have seen that the energy eigenvalue depends on the geometric quantum phase since there is the effective orbital quantum number, that is, $l \to l'=\Big(l-\frac{e\,\Phi_{AB}}{2\,\pi}\Big)$. Thus, the energy eigenvalue is a periodic function of the geometric phase, that is, $E_{n,l}(\Phi_{AB}\pm\nu\,\Phi_0)=E_{n,l\mp\nu}(\Phi_{AB})$ with $\nu=0,1,2,..$. This dependence of the eigenvalue on the geometric quantum phase gives us an electromagnetic analogue of the Aharonov-Bohm effect \cite{YA,MP}.

In {\bf section 2}, we derived the radial wave equation of the particles with harmonic oscillator plus Mie-type potential under the influence of flux field in a point-like defect and arrived the biconfluent Heun equation form. We then solved this equation using a power series method and obtained the non-compact energy eigenvalue expression by the Eq. (\ref{bb16}) and radial wave function by (\ref{bb9}). We have seen that the topological defect parameter $\alpha$ explicitly present in the energy eigenvalue breaks the degeneracy of the energy levels and get them modified compared to the flat space results. As special cases, we have presented two individual energy levels $E_{1,l}, E_{2,l}$ and radial wave function $\psi_{1,l}, \psi_{2,l}$ for the radial mode $n=1,2$ and others are in the same way.

In {\bf section 3}, we utilized the above eigenvalue solution to some known diatomic molecular potential models. The first one is being the harmonic oscillator plus Kratzer potential ({\tt sub-section:3 (a)}) and we presented the energy eigenvalue expression by the Eq. (\ref{cc3}). The second one is the harmonic oscillator plus modified Kratzer ({\tt sub-section:3(a)}) and presented the energy eigenvalue expression by (\ref{dd3}). Finally, we considered the harmonic plus attractive Coulomb potential ({\tt sub-section:3(a)}) and presented the energy eigenvalue expression by the (\ref{ff2}). As special cases, we have evaluated the energy levels $E_{1,l}, E_{2,l}$ and radial wave functions $\psi_{1,l}, \psi_{2,l}$ of the particles. We have seen that the eigenvalue solutions of the particles with these combined potential gets modified by the topological defect of the point-like global monopole, and the quantum flux field compared to the flat space results obtained in the literature.

In {\bf section 4}, we have considered a general form of the potential $V (r)=\Big(\beta\,r^2+\beta_1\,r+\frac{\beta_{-1}}{r}+\frac{\beta_{-2}}{r^2}+V_0\Big)$ and solved the non-relativistic wave equation in the same space-time background. Following the previous procedure, we obtained the energy eigenvalue by the expression (\ref{22}) and radial wave function Eq. (\ref{23}). We have seen that the eigenvalue solution here gets modified compared to the result in Ref. \cite{RV} by inverse square potential ($\sim \frac{1}{r^2}$), and the quantum flux field $\Phi_{AB}$.

\section*{Data Availability Statement}

No new data are generated or analyzed in this paper.

\section*{Conflict of Interest}

There is no potential conflict of interests in this paper.

\section*{Funding Statement}

No fund has received for this manuscript.

\section*{Acknowledgement}

We sincerely acknowledged the anonymous referee(s) for valuable comments, and suggestions.

\end{document}